\documentclass[aps,prd,email,twocolumn,showkeys,preprintnumbers,amsmath,amssymb,nofootinbib]{revtex4-2}

\usepackage{color}
\usepackage{latexsym}
\usepackage{amsmath}
\usepackage{amssymb}
\usepackage{eufrak}
\usepackage{euscript}
\usepackage[utf8]{inputenc} 
\usepackage[english]{babel} 
\usepackage[normalem]{ulem} 
\usepackage{graphics}
\usepackage{float}
\usepackage{graphicx}
\usepackage{picture}
\usepackage{pstricks,pst-coil}
\usepackage{pst-all}
\usepackage{blindtext}
\usepackage{dsfont}

\newcommand{\be}{\begin{equation}}
\newcommand{\ee}{\end{equation}}
\newcommand{\bn}{\begin{eqnarray}}
\newcommand{\en}{\end{eqnarray}}

\def\[{\left\lbrack}
\def\]{\right\rbrack}

\def\({\left(}
\def\){\right)}
	
\def\MyItem[#1]#2{\item[{#1}]#2}

\begin{document}

\title{Dimensional reduction and thermodynamics of a non-local scalar field theory}

\author{S. Duque Cesar}
\email{samduque@ufrrj.br} 
\affiliation{Department of Physics, Federal Rural University of Rio de Janeiro, BR 465-07, 23890-971, Seropédica, RJ, Brazil.}

\author{M. J. Neves}
\email{mariojr@ufrrj.br} 
\affiliation{Department of Physics, Federal Rural University of Rio de Janeiro, BR 465-07, 23890-971, Seropédica, RJ, Brazil.}


\begin{abstract}

In this paper, we study the dimensional reduction of a non-local scalar field theory from $1+3$ to $1+2$ dimensions. The bulk theory is governed by an entire-function non-local form factor $e^{-\Box_E/\Lambda^2}$, where $\Lambda$ sets a fundamental high-energy scale. By constraining external classical sources to a bidimensional plane, we perform the exact dimensional reduction, derive the effective planar Green's function in terms of the complementary error function, and propose the corresponding effective non-local pseudo-differential planar Lagrangian. Using the static propagator, we determine the interaction energy between two point-like planar charges, proving that the non-local scale regularizes the short-distance singularity at the origin while smoothly recovering the standard Yukawa potential at long distances. Furthermore, we obtain the retarded, advanced, and Feynman Green's functions in coordinate space, demonstrating that macroscopic causality and tree-level unitarity are preserved. Finally, implementing the Matsubara formalism, we investigate the finite-temperature regime and evaluate the exact thermal free energy, pressure, entropy, and heat capacity. We show that the effective planar theory exhibits a fractional effective number of degrees of freedom ($g_{\rm eff}=1/2$ for $m \ll T \ll \Lambda$, crossing over to $g_{\rm eff}=1$ for $T \gg \Lambda$) originating from kinematic bulk leakage and that pressure, entropy, and heat capacity remain positive at all temperatures.
\end{abstract}


\maketitle

\newpage

\section{Introduction}

Higher-derivative field theories and non-local interactions have attracted continuous interest in theoretical physics, primarily due to their inherent regularization properties in the ultraviolet (UV) regime \cite{PaisUhlenbeck1950}. The introduction of non-trivial differential operators in the Lagrangian density modifies the propagators at high energies, which frequently softens or eliminates classical divergences that afflict strictly local theories. Indeed, non-local formulations have been extensively applied in several research areas, as quantum gravity \cite{Briscese}, cosmological models \cite{Calcagni}, the study of spontaneous symmetry breaking in scalar QED \cite{Gama} and the Higgs mechanism \cite{Modesto}. Notably, alternative approaches to realize ghost-free spontaneous symmetry breaking have also been successfully formulated through weakly non-local covariant deformations \cite{Nortier2023, Chattopadhyay2024}. More recently, these generalized non-local formulations have seen rapid development, providing new theoretical insights into UV finiteness, gauge-invariant regulators, and collider phenomenology \cite{Moffat2025, Ajamieh2024, Biswas2014}, as well as the behavior of quantum entanglement in non-local field theories \cite{Landry2023}.
\par
A natural laboratory of great relevance for the application of non-local interactions arises in the physics of planar systems ($1+2$D). The most paradigmatic example is Pseudo-Electrodynamics (or Strictly Planar Electrodynamics), in which current sources are confined to a two-dimensional membrane, but the gauge field mediating the interaction permeates the three-dimensional ($3+1$D) volume of space-time \cite{Marino1993}. This formalism has proven fundamental in the precise description of interactions in low-dimensional condensed matter systems, such as graphene \cite{Marino2015}, where the effective in-plane dynamics inherits a non-local structure from the \textit{bulk}. In a broader context, the interplay between extra dimensions and non-locality has been shown to induce fuzzy branes, illustrating how lower-dimensional boundary fields are inherently smeared by bulk interactions \cite{Nortier2023b}. The study of fractional operators in these reduced dimensions has also provided crucial insights into the unitarity of non-local models \cite{MarinoUnitarity}, the emergence of effective Yukawa interactions \cite{MarinoYukawa}, and the behavior of point-like charges under specific boundary conditions \cite{Borges}. Furthermore, the scope of non-local interactions has been actively extended to investigate spinor and superfield dynamics \cite{Belchior2026, Gama2026, Nascimento2025}, Lorentz-violating topological terms \cite{Nascimento2026}, kink solutions \cite{Andrade2024}, and the evaluation of one-loop effective potentials \cite{Briscese2015}, underscoring the versatility of these non-local differential structures.
\par

In this paper, we start with a non-local scalar model in $1+3$ dimensions, governed by a $\Lambda$-scale that regularizes the ultraviolet divergences through the factor $e^{-\Box_E/\Lambda^2}$, see the reference \cite{Biswas2014}. Our central objective is to perform the dimensional reduction of this model to the $1+2$D space-time in order to extract the main characteristics of quantum field theory (QFT) for the effective planar model. We show that the non-local planar model softens short-distance singularities, while preserving the causal integrity and unitarity of the model without introducing ghost states. Furthermore, we explore the finite-temperature regime to unveil how the planar confinement modulates the thermal properties. By establishing the exact thermodynamic behavior of the free effective theory, we provide the mathematical groundwork required for future non-perturbative investigations of interacting planar models, where genuine phase transitions and dynamical symmetry breaking phenomena can take place.
\par

The paper is organized as follows: In Section II, we present the non-local scalar model in Euclidean space and carry out the dimensional reduction to $1+2$ dimensions, obtaining the planar Green's function and proposing the effective planar Lagrangian. In Section III, we compute the static interaction energy between two point-like charges through the static propagator. In Section IV, we investigate the causality of the theory by calculating the retarded, advanced, and Feynman Green's functions in coordinate space. In Section V, we show unitarity of the planar model at tree level. In Section VI, we employ the Matsubara formalism to evaluate the exact thermal free energy, discussing the fractional degrees of freedom and the unconditional thermodynamic stability of the reduced planar bath. Finally, in Section VII, we present our conclusions.
\vspace{1pt}

We adopt the natural units of $k_{B}=\hbar=c=1$ in which the Minkowski metric has the signature $\eta^{\mu\nu}=(1,-1,-1,-1)$.
In $1+2$ dimensions, we use the greek bar index $\bar{\mu},\bar{\nu}=\left\{ \, 0, \, 1, \, 2 \, \right\}$ with the metric $\eta^{\bar{\mu}\bar{\nu}}=(1,-1,-1)$.
\section{Dimensional reduction in the non-local scalar model}
\label{sec2}
The non-local scalar field model is set by the action defined in the Euclidian space :
\begin{equation}\label{Lscalar}
S[\phi]=\int d^{4}x_{E} \left[ \, \frac{1}{2} \, \phi(x_{E}) \, \hat{{\cal O}}(-\Box_{E}) \phi(x_{E}) + J(x_{E}) \, \phi(x_{E}) \, \right] \, ,   
\end{equation}
where the $\hat{{\cal O}}(-\Box_{E})$-operator is 

\begin{eqnarray}
\hat{{\cal O}}(-\Box_{E})=\left(\,-\Box_{E}+m^2\,\right) \, e^{-\frac{\Box_{E}}{\Lambda^2}} \; ,
\end{eqnarray}
in which $\Box_{E} = \partial_4^2 + \partial_x^2 + \partial_y^2+\partial_{z}^2$ is the D'Alembertian operator, 
$m$ is the mass of the scalar field, $\Lambda$ is the non-local scale with mass dimension, and $J$ is a classical source. 
The usual scalar theory in Euclidian space is recovered in the limit $\Lambda \rightarrow \infty$. Since $\Lambda$ is a high energy scale, 
in some points of this work, we can consider the condition $\Lambda \gg m$. 
%
%
%
%
Using the action principle in (\ref{Lscalar}), the non-local Klein-Gordon (KG) equation is
\begin{eqnarray}\label{Eqphi}
\left(\,-\Box_{E}+m^2\,\right) \, e^{-\frac{\Box_{E}}{\Lambda^2}}\phi(x_{E})=J(x_{E}) \; .   
\end{eqnarray}
The general solution to eq. (\ref{Eqphi}) is
\begin{eqnarray}\label{solphi}
\phi(x_{E})=\phi_{0}(x_{E})+\int d^{4}x_{E}^{\prime} \, \Delta(x_{E}-x_{E}^{\prime}) \, J(x_{E}^{\prime}) \; ,   
\end{eqnarray}
where $\phi_{0}(x_{E})$ is the solution for the free KG equation, and $\Delta(x_{E}-x_{E}^{\prime})$ is the Green's function of the non-local operator 
from (\ref{Eqphi}) that satisfies the equation :
\begin{equation}
\left(\,-\Box_{E}+m^2\,\right) \, e^{-\frac{\Box_{E}}{\Lambda^2}}\Delta(x_{E}-x_{E}^{\prime})=\delta^{4}(x_E-x_{E}^{\prime}) \, .  
\end{equation}
Using the Fourier transform for the Green function, 
we obtain the integral 
\begin{equation}\label{DFG}
\Delta(x_{E}-x_{E}^{\prime}) = \int \frac{d^4p_{E}}{(2\pi)^4} \frac{e^{-p_{E}^2/\Lambda^2}}{p_{E}^2+m^2} \; e^{-i\,p_{E}\cdot(x_{E}-x_{E}^{\prime})} \; .
\end{equation}
Notice that in the momentum space, the Green function is null in the ultraviolet regime.

To obtain the effective theory in $1+2$ dimensions, we constraint the dynamics of the scalar model confined to a two-dimensional plane. The planar condition is such that the sources satisfy the relation: 
\begin{eqnarray}\label{J}
J(x_{E})=j(\bar{x}_{E}) \, \delta(z) \; ,   
\end{eqnarray}
where $\bar{x}_{E}=\{ \, x_4 \, , \, x \, , \, y \, \}$ are the coordinates in the $3$-Euclidian space, and $j(\bar{x}_{E})$ is the new source 
defined on the bidimensional plane. Substituting this condition in the solution (\ref{solphi}), the $z^{\prime}$-integration yields 
\begin{eqnarray}\label{solphi3D}
\phi(\bar{x}_{E})=\phi_{0}(\bar{x}_{E})+\int d^{3}x_{E}^{\prime} \, \Delta_{1+2}(\bar{x}_{E}-\bar{x}_{E}^{\prime}) \, j(\bar{x}_{E}^{\prime}) \; , 
\end{eqnarray}
where the Green function in $3$-dimensions is given by $\Delta_{1+2}(\bar{x}_{E}-\bar{x}_{E}^{\prime})=\Delta(x_{E}-x_{E}^{\prime})|_{z=z^{\prime}=0}$. Therefore, we write
\begin{eqnarray}\label{Delta1+2}
\Delta_{1+2}(\bar{x}_{E}-\bar{x}_{E}^{\prime}) &=& \int \frac{d^3\bar{p}_{E}}{(2\pi)^3} \, e^{-i\bar{p}_{E}\cdot(\bar{x}_{E}-\bar{x}_{E}^{\prime})} 
\, e^{-\bar{p}_{E}^2/\Lambda^2} 
\nonumber \\
&&
\hspace{-0.4cm}
\times \int_{-\infty}^{\infty} \frac{dp_z}{2\pi} \, \frac{e^{-p_{z}^2/\Lambda^2}}{p_z^2 + \bar{p}_{E}^2 + m^2} \; ,
\end{eqnarray}
in which the bar momentum contains the components $p_{E}^{\bar{\mu}} = (p_4, p_x, p_y)$, 
with $\bar{p}_{E}^2=p_{4}^2+p_{x}^2+p_y^2>0$. The result of (\ref{Delta1+2}) is 
\begin{eqnarray}\label{Delta1+2result}
\Delta_{1+2}(\bar{x}_{E}-\bar{x}_{E}^{\prime}) = \int \frac{d^3\bar{p}_{E}}{(2\pi)^3} \, e^{-i\bar{p}_{E}\cdot(\bar{x}_{E}-\bar{x}_{E}^{\prime})} 
\nonumber \\
\times \, e^{m^2/\Lambda^2} \, \frac{\mbox{erfc}(\sqrt{\bar{p}_{E}^2+m^2}/\Lambda)}{2\,\sqrt{\bar{p}_{E}^2+m^2}} \; ,
\end{eqnarray}
%
%
%
%
where $\mbox{erfc}$ is a complementary error function. 
In the ultraviolet regime, the scalar propagator runs as $\sim e^{-\bar{p}_{E}^2/\Lambda^2}/\bar{p}_{E}^{2}$, that is null in the limit $\bar{p}^2_{E} \rightarrow \infty$. 

Our task now is to obtain the lagrangian density that leads to the Green function (\ref{Delta1+2result}).  
We propose the lagrangian in the Euclidian space 
\begin{equation}\label{L3D}
\mathcal{L}_{NLKG}^{1+2} = 
\frac{1}{2} \, \phi(x_{E}) \, \hat{{\cal O}}(-\bar{\Box}_{E}) \, \hat{W}(-\bar{\Box}_{E}) \, \phi(x_E)
\; ,
\end{equation}
where 
$\hat{W}(\bar{\Box}_{E})$ is a non-local operator that is function of the D'Alembertian operator $\bar{\Box}_{E}=\partial_{4}^2+\partial_{x}^2+\partial_{y}^2$ 
in 3D Euclidian space. 
%
%
%
The non-local operator $\hat{{\cal A}}(-\bar{\Box}_{E})=\hat{{\cal O}}(-\bar{\Box}_{E}) \, \hat{W}(-\bar{\Box}_{E})$ 
in the momentum space $\bar{\Box}_{E} \rightarrow -\bar{p}_{E}^2$ is given by
\begin{equation}\label{Ok}
{\cal A}(\bar{p}_{E}^2) = \left( \, \bar{p}_{E}^2 + m^2 \, \right) \, e^{\bar{p}_{E}^2/\Lambda^2} \, W(\bar{p}_{E}^2)
\; .
\end{equation}
The Green function in the momentum space is defined by the inverse of the function $\mathcal{A}(\bar{p}_{E}^2)$. Comparing (\ref{Ok}) with (\ref{Delta1+2result}), 
we have the relation 
\begin{eqnarray}
{\cal A}(\bar{p}_{E}^2)^{-1}=e^{m^2/\Lambda^2} \, \frac{\mbox{erfc}\left(\sqrt{\bar{p}_{E}^2+m^2}/\Lambda\right)}{2\,\sqrt{\bar{p}_{E}^2+m^2}} \; ,
\end{eqnarray}
that yields the result for the $W(\bar{p}_{E}^2)$-function
\begin{equation}
W(\bar{p}_{E}^2) = \frac{2}{\sqrt{\bar{p}_{E}^2 + m^2}} \frac{e^{-\left(\bar{p}_{E}^2+m^2\right)/\Lambda^2}}{\mbox{erfc}\left(\sqrt{\bar{p}_{E}^2+m^2}/\Lambda\right)} \; .
\end{equation}
In the momentum space, the non-local scalar model in $1+2$ dimensions
\begin{equation}
\mathcal{L}_{NLKG}^{1+2} = \frac{1}{2} \, \phi(x_{E}) \, \frac{2\,\sqrt{-\bar{\Box}_{E}+m^2}}{\mbox{erfc}\left(\sqrt{-\bar{\Box}_{E}+m^2}/\Lambda\right)} \, \phi(x_{E}) \, ,
\end{equation}
or in the standard form
\begin{equation}\label{L3DW}
\mathcal{L}_{NLKG}^{1+2} = \frac{1}{2} \, \partial_{E\bar{\mu}} \phi \, \hat{{\cal B}}(-\bar{\Box}_{E}) \, \partial_{E}^{\bar{\mu}}\phi
+\frac{1}{2} \, m^2 \, \phi \, \hat{{\cal B}}(-\bar{\Box}_{E})  \, \phi \; ,
\end{equation}
where 
\begin{equation}
\hat{{\cal B}}(-\bar{\Box}_{E})=\frac{2}{\sqrt{-\bar{\Box}_{E}+m^2}} \, \frac{1}{ \mbox{erfc}\left(\sqrt{-\bar{\Box}_{E}+m^2}/\Lambda\right) }  \; .
\end{equation}
By consistence, the limit $\Lambda \rightarrow \infty$ recovers the pseudo-scalar model obtained in the ref. \cite{MarinoYukawa}. 

Similarly to the scalar case, the same approach can be applied for the non-local electrodynamics in Euclidian space : 
\begin{eqnarray}\label{LEM}
{\cal L}_{NLED} &=& \frac{1}{4} \, F_{\mu\nu} \, \hat{{\cal O}}(-\Box_{E}) \,  F^{\mu\nu}
\nonumber \\
&&
\hspace{-0.5cm}
+\frac{1}{2\,\xi} \, (\partial_{\mu}A^{\mu}) \, \hat{{\cal O}}(-\Box_{E}) \,(\partial_{\nu}A^{\nu}) -J_{\mu}A^{\mu} \, ,   
\end{eqnarray}
in which $J^{\mu}$ is an external source, with the strength field tensor $F_{\mu\nu}=\partial_{\mu}A_{\nu}-\partial_{\nu}A_{\mu}$, and 
$\xi$ is a gauge fixing parameter. The result for the dimensional reduction is the non-local pseudo-electrodynamics :
 \begin{eqnarray}\label{LPseudoNLEM}
{\cal L}_{NLED}^{1+2}=\frac{1}{4} \, F_{\bar{\mu}\bar{\nu}} \, \hat{{\cal B}}(-\bar{\Box}_{E}) \,  F^{\bar{\mu}\bar{\nu}}-J_{\bar{\mu}}\,A^{\bar{\mu}} \; ,   
\end{eqnarray}
where $F_{\bar{\mu}\bar{\nu}}=(i\,E_{x},i\,E_{y},-B)$ is the EM strength field tensor in the $3D$ Euclidian space.

We will study the static energy between two particles in the next section.

\section{Static energy between two scalar particles}

We consider the static case of two scalar particles at rest whose interaction is mediated by the energy associated with the propagator of the planar model (\ref{L3DW}) in which $p_{0}=0$. The Green function in the momentum space in this case is
\begin{equation}\label{eq:prop_estatico}
D_{1+2}(\mathbf{p}) = e^{m^2/\Lambda^2} \, \frac{\mbox{erfc}(\sqrt{\mathbf{p}^2+m^2}/\Lambda)}{2\,\sqrt{\mathbf{p}^2+m^2}} \; .
\end{equation}

The static energy between two point-like charges $\sigma_1$ and $\sigma_2$ separated by a radial distance $r = |\mathbf{r}|$ in the plane is given by
\begin{equation}
U(r) = - \, \sigma_1 \, \sigma_2 \int \frac{d^2\mathbf{p}}{(2\pi)^2} \, e^{i \, \mathbf{p} \cdot \mathbf{r}} \, D_{1+2}(\mathbf{p}) \; .
\end{equation}
Using the polar coordinates in momentum space, we obtain 
\begin{equation}
U(r) = -\frac{\sigma_1 \, \sigma_2}{2} \, e^{m^2/\Lambda^2} \! \int_0^\infty \frac{dp}{2\pi} \, p \, J_0(pr) 
\frac{\mbox{erfc}(\sqrt{p^2+m^2}/\Lambda)}{\sqrt{p^2+m^2}} \; ,
\end{equation}
where $J_{0}$ is a Bessel function, and the result of the integrals
yields the static energy
\begin{equation}
\begin{split}
U(r) &= -\frac{\sigma_1  \, \sigma_2}{8\pi r} \, e^{m^2/\Lambda^2} \\
&\quad \times \left[ \, \mbox{erfc}\left( \frac{m}{\Lambda} - \frac{\Lambda\,r}{2} \right) e^{-m \, r} 
- \mbox{erfc} \left( \frac{m}{\Lambda} + \frac{\Lambda\,r}{2} \right) e^{m \, r} \, \right] \; .
\end{split}
\end{equation}
In the local limit $\Lambda \gg m$, using $\mbox{erfc}(m/\Lambda - \Lambda r/2) \to 2$ and $\mbox{erfc}(m/\Lambda + \Lambda r/2) \to 0$ for any $r > 0$, the static energy smoothly recovers the standard Yukawa potential :
\begin{equation}
\lim_{\Lambda \to \infty} U(r) = -\frac{\sigma_1 \, \sigma_2}{4\pi r} \, e^{-mr} \; .
\end{equation}
Crucially, at short distances ($r \to 0$), the non-local form factor cures the classical electrostatic divergence. Expanding the complementary error functions and the exponentials around $r=0$, the $1/r$ singularity cancels out exactly, yielding a finite value at the origin:
\begin{equation}\label{U0}
U(0) = -\frac{\sigma_1 \sigma_2 \Lambda}{4\pi^{3/2}} + \frac{\sigma_1 \sigma_2 m}{4\pi} \, e^{m^2/\Lambda^2} \, \mbox{erfc}\left( \frac{m}{\Lambda} \right) \; .
\end{equation}
This demonstrates that the non-locality scale $\Lambda$ serves as an intrinsic physical ultraviolet regulator that renders the self-energy and the interaction energy at the coincidence limit strictly finite.

\begin{figure}[htbp]
    \centering
    \includegraphics[width=1\columnwidth]{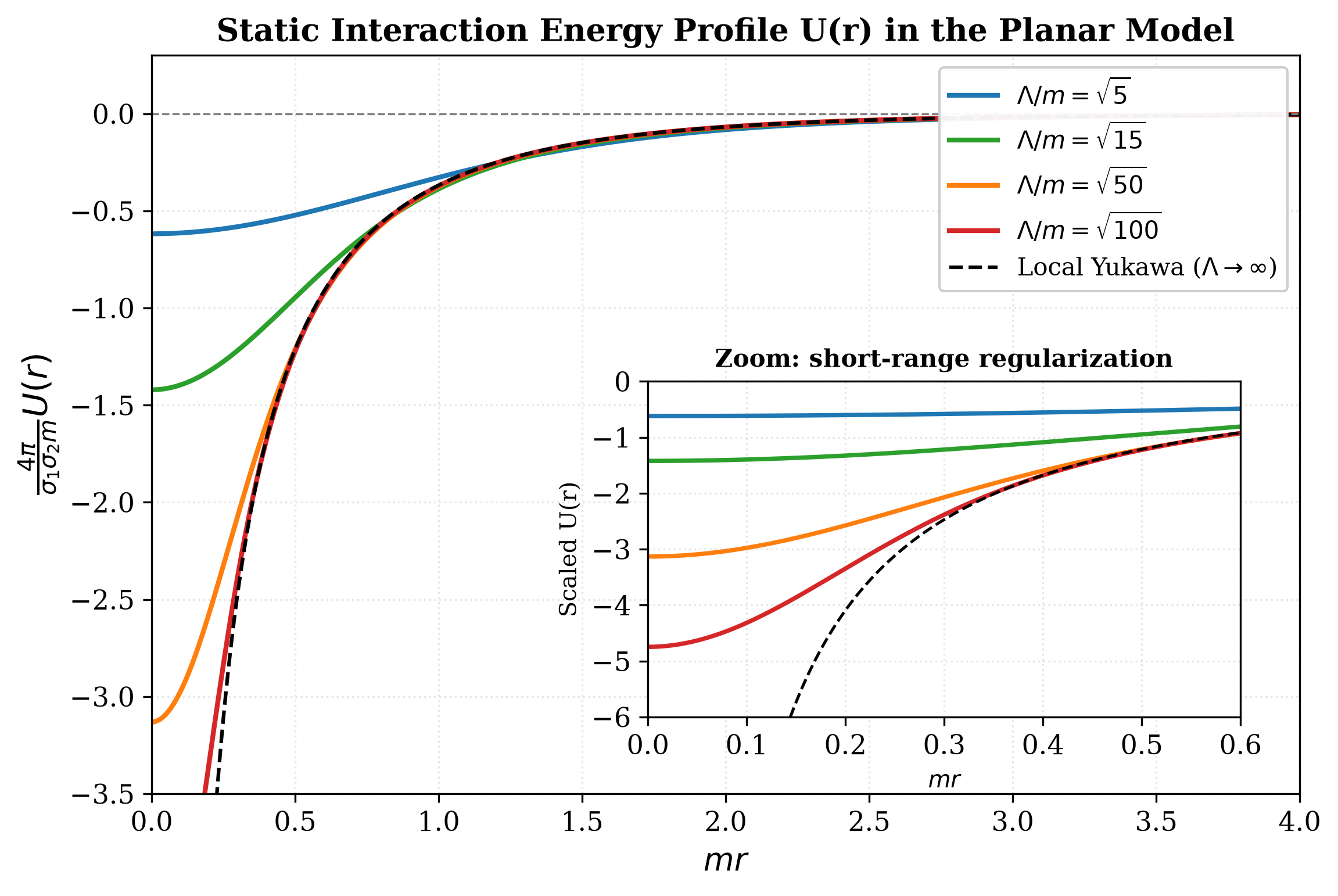}
    \caption{The static interaction energy $U(r)$ as a function of the dimensionless variable $mr$. We plot the profiles for $\Lambda/m=\sqrt{5}$, $\Lambda/m=\sqrt{15}$, $\Lambda/m=\sqrt{50}$, and $\Lambda/m=\sqrt{100}$, compared to the standard local Yukawa potential ($\Lambda \to \infty$). The inset highlights the short-range behavior ($mr \to 0$), showing the smooth regularization of the singularity at the origin.}
    \label{fig:energia_potencial}
\end{figure}

The dimensionless static energy $(4\pi/(\sigma_1\sigma_2 m)) U(r)$ as a function of the dimensionless distance $mr$ is illustrated in Figure \ref{fig:energia_potencial} for $\Lambda/m = \sqrt{5}$, $\sqrt{15}$, $\sqrt{50}$, and $\sqrt{100}$. As $\Lambda/m$ increases, the potential rapidly approaches the local Yukawa curve, while the inset highlights the short-distance regularization at $r \to 0$.

\section{Causality in the planar non-local model}
\label{sec5}

\subsection{Retarded and advanced Green functions}

In this section, we investigate the macroscopic causality of the planar scalar model (\ref{L3DW}). 
Therefore, we evaluate the retarded and advanced Green's functions in Minkowski spacetime in $1+2$ dimensions. 
These Green functions are obtained through the standard prescriptions : $p_0 \rightarrow p_0 + i\epsilon$ for the 
retarded function $D_{1+2}^{(-)}$, and $p_0 \rightarrow p_0 - i\epsilon$ for the advanced function $D_{1+2}^{(+)}$. Thus, we have
\begin{equation}\label{Dpm}
D_{1+2}^{(\pm)}(p_0,{\bf p}) = e^{m^2/\Lambda^2} \frac{\mbox{erfc}\left(\sqrt{-(p_{0}\mp i\,\epsilon)^2 +{\bf p}^2 + m^2}/\Lambda\right) }{2 \,\sqrt{-(p_{0}\mp i\,\epsilon)^2 +{\bf p}^2 + m^2}} \; .
\end{equation}
In coordinate space, the retarded and advanced Green functions are defined by the Fourier transform :
\begin{equation}\label{DeltaFT}
\Delta_{1+2}^{(\mp)}(\bar{x}) = \int \frac{d^3\bar{p}}{(2\pi)^3} \, e^{-i\bar{p}\cdot\bar{x}} \, D_{1+2}^{(\pm)}(\bar{p}) \; ,
\end{equation}
where $\bar{x} = (\tau, {\bf x})$ with $\tau = t - t^\prime$, $r = |{\bf x}-{\bf x}^\prime|$, and $\bar{x}^2 = \tau^2 - r^2$. 

To evaluate this integral, we analyze the analytic structure of the momentum-space propagator (\ref{Dpm}). Using the identity $\mbox{erfc}(z) = 1 - \mbox{erf}(z)$ and the power series representation of the error function,
\begin{equation}
\frac{\mbox{erf}(z)}{z} = \frac{2}{\sqrt{\pi}} \sum_{k=0}^{\infty} \frac{(-1)^k \, z^{2k}}{k! \, (2k+1)} \; ,
\end{equation}
we observe that $\mbox{erf}(\sqrt{\Omega^2}/\Lambda)/\sqrt{\Omega^2}$ is an entire function of the complex variable $\Omega^2 = -(p_0 \mp i\epsilon)^2 + {\bf p}^2 + m^2$. An entire function has no poles and no branch cuts in the entire finite complex plane. Consequently, the discontinuity across the real axis (the spectral imaginary part) is determined solely by the branch cut of the square root denominator:
\begin{eqnarray}\label{DiscD}
\mbox{Disc}\left[ D_{1+2}(\bar{p}^2) \right] &=& 2i \, \mbox{Im}\left[ D_{1+2}(\bar{p}^2) \right] 
\nonumber \\
&=& i \, \frac{e^{m^2/\Lambda^2}}{\sqrt{\bar{p}^2 - m^2}} \, \Theta(\bar{p}^2 - m^2) \; .
\end{eqnarray}
In field theory, the causal Green functions are directly related to the field commutator:
\begin{equation}
\Delta_{1+2}^{(\mp)}(\bar{x}) = \pm \, \Theta(\pm\tau) \, \langle 0 | \left[ \phi(\bar{x}), \phi(0) \right] | 0 \rangle \; .
\end{equation}
The commutator is determined entirely by the spectral discontinuity (\ref{DiscD}) :
\begin{eqnarray}\label{Commutator}
\langle 0 | \left[ \phi(\bar{x}), \phi(0) \right] | 0 \rangle = e^{m^2/\Lambda^2} \int \frac{d^3\bar{p}}{(2\pi)^3} \, e^{-i\bar{p}\cdot\bar{x}} \, 
\nonumber \\
\times \frac{\mbox{sgn}(p_0)}{\sqrt{\bar{p}^2 - m^2}} \, \Theta(\bar{p}^2 - m^2) 
= \frac{i \, e^{m^2/\Lambda^2}}{2\pi} \, \mbox{sgn}(\tau) 
\nonumber \\
\times \left[ \, \delta(\bar{x}^2) - \Theta(\bar{x}^2) \, \frac{m \, J_1(m\sqrt{\bar{x}^2})}{2\sqrt{\bar{x}^2}} \, \right] \; .
\end{eqnarray}
Multiplying by $\pm \Theta(\pm\tau)$, we obtain the retarded and advanced Green functions in coordinate space:
\begin{equation}\label{Delta1+2resultcoord}
\Delta_{1+2}^{(\mp)}(\bar{x}) = \frac{i\,\Theta(\pm \tau) \, e^{m^2/\Lambda^2}}{2\pi} \left[ \, \delta(\bar{x}^2) - \Theta(\bar{x}^2)\, \frac{m\, J_1(m\,\sqrt{\bar{x}^2})}{2\,\sqrt{\bar{x}^2}} \, \right] \; .
\end{equation}
The presence of $\Theta(\pm\tau)$ and $\Theta(\bar{x}^2)$ explicitly proves that:
\begin{enumerate}
    \item The retarded Green function $\Delta_{1+2}^{(-)}(\bar{x})$ is identically zero for $\tau < 0$, strictly preserving macroscopic temporal causality.
    \item Both Green functions strictly vanish for space-like intervals, $\bar{x}^2 = \tau^2 - r^2 < 0$, guaranteeing that no signal propagates outside the light-cone.
    \item In the local limit $\Lambda \gg m$, the prefactor $e^{m^2/\Lambda^2} \to 1$, smoothly reproducing the causal Green function of the local massive scalar theory in $1+2$ dimensions.
\end{enumerate}

\subsection{Feynman Green's function in coordinate space}

To evaluate the Feynman Green's function in coordinate space, we implement the Feynman prescription $\bar{p}^2 \rightarrow \bar{p}^2 + i\epsilon$, which sets the time-ordered propagator :
\begin{equation}\label{DFdef}
\Delta_{1+2}^{(F)}(\bar{x}) = \int \frac{d^3\bar{p}}{(2\pi)^3} \, e^{-i\,\bar{p}\cdot\bar{x}} \, \frac{e^{m^2/\Lambda^2} \, \mbox{erfc}\left(\sqrt{-\bar{p}^2+m^2-i\epsilon}/\Lambda\right)}{2\,\sqrt{-\bar{p}^2+m^2-i\epsilon}} \; .
\end{equation}
This integral is most conveniently computed by first evaluating the corresponding Euclidean Green function $\Delta_{1+2}^E(\bar{x}_E)$ and then performing an analytic continuation back to Minkowski spacetime. Using the proper-time Schwinger representation:
\begin{equation}
\frac{\mbox{erfc}\left(\sqrt{\bar{p}_E^2+m^2}/\Lambda\right)}{2\,\sqrt{\bar{p}_E^2+m^2}} = \int_{1/\Lambda^2}^\infty \frac{ds}{2\sqrt{\pi s}} \, e^{-s(\bar{p}_E^2+m^2)} \; ,
\end{equation}
the Euclidean Green function in coordinate space is expressed as :
\begin{eqnarray}\label{DeltaEcoord}
&&
\Delta_{1+2}^E(\bar{x}_E) = e^{m^2/\Lambda^2} \int_{1/\Lambda^2}^\infty \frac{ds}{2\sqrt{\pi s}} \, e^{-s m^2} 
\nonumber \\
&&
\times \int \frac{d^3\bar{p}_E}{(2\pi)^3} \, e^{i\bar{p}_E\cdot\bar{x}_E - s \bar{p}_E^2}
\nonumber \\
&=& \frac{e^{m^2/\Lambda^2}}{16\pi^2} \int_{1/\Lambda^2}^\infty \frac{ds}{s^2} \, \exp\left( -\frac{\bar{x}_E^2}{4s} - s m^2 \right) \; ,
\end{eqnarray}
where $\bar{x}_E^2 = x_4^2 + {\bf x}^2 > 0$ is the Euclidean distance squared. 

Analytically, continuing this result back to Minkowski spacetime through the Wick rotation $\bar{x}_E^2 \rightarrow -\bar{x}^2 + i\epsilon$ and multiplying by the global factor $+i$, the Feynman Green function in coordinate space is obtained in the exact closed integral form:
\begin{equation}\label{DeltaFresult}
\Delta_{1+2}^{(F)}(\bar{x}) = \frac{i \, e^{m^2/\Lambda^2}}{16\pi^2} \int_{1/\Lambda^2}^\infty \frac{ds}{s^2} \, \exp\left( \frac{\bar{x}^2 - i\epsilon}{4s} - s m^2 \right) \; .
\end{equation}
In the local limit $\Lambda \rightarrow \infty$, the lower integration limit extends to $s=0$, and using the standard integral representation of the Macdonald function,
\begin{equation}
\int_0^\infty \frac{ds}{s^2} \, \exp\left( -\frac{a}{4s} - s b \right) = \frac{4\sqrt{b}}{\sqrt{a}} \, K_1\left(\sqrt{ab}\right) \; ,
\end{equation}
Eq. (\ref{DeltaFresult}) smoothly reduces to the standard Feynman propagator of the local pseudo-scalar theory:
\begin{equation}\label{DeltaFlocal}
\lim_{\Lambda \to \infty} \Delta_{1+2}^{(F)}(\bar{x}) = \frac{i\,m}{4\pi^2 \sqrt{-\bar{x}^2 + i\epsilon}} \, K_1\left( m\sqrt{-\bar{x}^2 + i\epsilon} \right) \; .
\end{equation}
At finite $\Lambda$, at the coincidence limit on the light-cone ($\bar{x}^2 \to 0$), the integral in (\ref{DeltaFresult}) yields a strictly finite value:
\begin{equation}\label{DeltaF0}
\Delta_{1+2}^{(F)}(0) = \frac{i\,\Lambda^2}{16\pi^2} \left[ 1 - \frac{m^2}{\Lambda^2} \, e^{m^2/\Lambda^2} \, E_1\left(\frac{m^2}{\Lambda^2}\right) \right],
\end{equation}
where $E_1(z) = \int_z^\infty \frac{e^{-t}}{t}\,dt = -\mbox{Ei}(-z)$ is the exponential integral. This explicitly demonstrates that the non-local scale $\Lambda$ acts as a genuine ultraviolet regulator that removes the singularity of the Feynman propagator at short distances while maintaining the full causal structure of the theory.
\section{Unitarity at tree level}

In QFT, unitarity is one of the main required properties for a consistent theory. The unitarity condition imposes that the $\hat{S}$-operator satisfies to $\hat{S}^{\dagger}\,\hat{S}=1$, where the elements of $S$-matrix are defined by
\begin{eqnarray}
S_{fi}=\lim_{t_{0} \rightarrow -\infty}\lim_{t \rightarrow +\infty}\langle f | e^{-i\hat{H}(t-t_{0})} | i \rangle \; ,
\end{eqnarray}
for a system set by the hamiltonian $\hat{H}$-operator, in which $|i\rangle$ and $|f\rangle$ are the free asymptotic states. The $S$-operator can be written as 
$S=\mathds{1}+i\,T$, in which the unitarity condition leads to relation 
$i[T^{\dagger}-T]=T^{\dagger}\,T$. Defining the element $T_{ii}=\langle i | T | i \rangle$, it satisfies the relation   
\begin{eqnarray}
2\,\mbox{Im}(T_{ii})=\sum_{f}T_{if}^{\dagger}\,T_{fi}=\sum_{f}T_{fi}^{\ast}\,T_{fi} \; ,
\end{eqnarray}
that is known as the Optical theorem. We can relate $T_{ii}$ with the Feynman propagator by $T_{ii}=(2\pi)^3\,\delta^3(0)\,\Delta_{F}(\bar{x}-\bar{x}^{\prime})$, where the unitarity condition leads to expression
\begin{equation}\label{RelDeltaF}
\begin{split}
\Delta_{F}^{\ast}(\bar{x})-\Delta_{F}(\bar{x}) &= -i\int d\Phi \, (2\pi)^3\,\delta^{3}(0) \\
&\quad \times \int\frac{d^3\bar{x}^{\prime}}{(2\pi)^3} \, \Delta_{F}^{\ast}(\bar{x}^{\prime}) \,\Delta_{F}(\bar{x}-\bar{x}^{\prime}) \; ,
\end{split}
\end{equation}
where $d\Phi$ is a phase factor that when integrated to Dirac delta function yields the characteristic time of the system
\begin{eqnarray}
\int d\Phi \, (2\pi)^3\,\delta^{3}(0)={\cal T}^{-1} \; .
\end{eqnarray} 
The Fourier transform in (\ref{RelDeltaF}) leads to relation of Feynman Green function in the momentum space :  
\begin{eqnarray}\label{GFrel}
D_{F}^{\ast}(\bar{k}^2)-D_{F}(\bar{k}^2)=-\,i \, {\cal T}^{-1} \, D_{F}^{\ast}(\bar{k}^2)\,D_{F}(\bar{k}^2) \; .
\end{eqnarray}
In the non-local model reduced to $1+2$ dimensions, the Feynman Green function in the momentum space is
\begin{equation}
D_{F}(\bar{k}^2) \simeq 
\frac{\mbox{erfc}(\sqrt{ -\bar{k}^2+m^2 }/\Lambda) }{2 \, \sqrt{-\bar{k}^2 + m^2-i\,\epsilon}} \; ,
\end{equation}
with $\epsilon > 0$. Substituting in the relation (\ref{GFrel}), we obtain 
\begin{eqnarray}\label{Imrel}
\frac{\Im[\,\sqrt{-\bar{k}^2+m^2-i\,\epsilon}\,]}{[ \, (\bar{k}^2-m^2)^2+\epsilon^2 \, ]^{1/2}} = \frac{-\,{\cal T}^{-1}/4}{[\, (\bar{k}^2-m^2)^2+\epsilon^2 \, ]^{1/2}} \; .   
\end{eqnarray}
Squaring (\ref{Imrel}) and multiplying in both sides by $\epsilon$, the limit $\epsilon \rightarrow 0$ yields the relation
\begin{eqnarray}
\left. \frac{{\cal T}^{-2}}{16}=\Im\left[\,\sqrt{-\bar{k}^2+m^2-i\,\epsilon}\,\right]^2 \right|_{k^2=m^2} \; ,   
\end{eqnarray}
whose result for ${\cal T}$ is
\begin{eqnarray}
{\cal T}^{-1} = 2 \, \sqrt{2 \, \epsilon} \; .   
\end{eqnarray}
This result shows that the non-planar model is unitary at tree level.

\section{Thermodynamics of the non-Local planar model}
\label{sec_thermo}

To expand the physical scope of the effective theory, we now evaluate its thermodynamic behavior at finite temperature $T = \beta^{-1}$, with $k_{B}=1$ in natural units. We employ the Matsubara formalism, performing a Wick rotation to imaginary time $t \to -i\tau$, with $\tau \in [0, \beta]$, and periodic boundary conditions for the scalar field, as usual. The continuous energy spectrum is replaced by the discrete Matsubara frequencies $\omega_n = 2\pi n \, T$, with $n$ integer, and the Euclidean momentum squared becomes $p_E^2 = \omega_n^2 + \mathbf{p}^2$.
The thermal properties are described by the partition function $\mathcal{Z}$, defined by the path integral:
%
\begin{eqnarray}
    \mathcal{Z} &=& \int \mathcal{D}\phi \, \exp\left( - \frac{1}{2} \int_0^\beta d\tau \int d^2\mathbf{x} \, \phi \, \mathcal{O}_E \, \phi \right) 
    \nonumber \\
    &&
    = \left[ \det(\mathcal{O}_E) \right]^{-1/2} \; .
\end{eqnarray}
Transforming to Euclidean space, the exact kinetic operator $\mathcal{O}_E$ derived in Section \ref{sec2} assumes a strictly positive argument. Factoring out irrelevant overall constants that do not contribute to the thermal dynamics, the operator reads :
\begin{equation}
\mathcal{O}_E(\omega_n, \mathbf{p}) = 
\frac{2\,\sqrt{\omega_{n}^2+{\bf p}^2+m^2}}{\mbox{erfc}\left(\sqrt{\omega_{n}^2+{\bf p}^2+m^2}/\Lambda\right)} \; .
\end{equation}

The free energy density $\mathcal{F} = -\frac{1}{\beta V} \ln \mathcal{Z}$ is evaluated using the identity $\ln(\det \mathcal{O}_E) = \text{Tr}(\ln \mathcal{O}_E)$, which translates into a sum over the Matsubara frequencies and a spatial momentum integral :
\begin{equation}
\mathcal{F} = \frac{T}{2} \sum_{n=-\infty}^{\infty} \int \frac{d^2\mathbf{p}}{(2\pi)^2} \ln \left[ \, \mathcal{O}_E(\omega_n, \mathbf{p}) \, \right] \; .
\end{equation}

The kinetic operator $\mathcal{O}_E$ in the Euclidean momentum space reads:
\begin{equation}\label{OEexact}
\mathcal{O}_E(\omega_n, \mathbf{p}) = \frac{2\,\sqrt{\omega_{n}^2+{\bf p}^2+m^2}}{\mbox{erfc}\left(\sqrt{\omega_{n}^2+{\bf p}^2+m^2}/\Lambda\right)} \; .
\end{equation}
The free energy density $\mathcal{F} = -\frac{1}{\beta V} \ln \mathcal{Z}$ is evaluated using the identity $\ln(\det \mathcal{O}_E) = \text{Tr}(\ln \mathcal{O}_E)$, which translates into a sum over the discrete Matsubara frequencies and an integral over the spatial planar momentum:
\begin{equation}\label{Fdet}
\mathcal{F} = \frac{T}{2} \sum_{n=-\infty}^{\infty} \int \frac{d^2\mathbf{p}}{(2\pi)^2} \ln \left[ \, \mathcal{O}_E(\omega_n, \mathbf{p}) \, \right] \; .
\end{equation}
Taking the logarithm of the kinetic operator (\ref{OEexact}), we obtain:
\begin{eqnarray}\label{lnOE}
\ln \mathcal{O}_E &=& \ln 2 + \frac{1}{2} \, \ln \left(\omega_n^2 + \mathbf{p}^2 + m^2\right) 
\nonumber \\
&& 
- \ln\left[ \mbox{erfc} \left(\frac{\sqrt{\omega_n^2 + \mathbf{p}^2 + m^2}}{\Lambda}\right) \right] \; .
\end{eqnarray}
Dropping the irrelevant constant $\ln 2$ which does not contribute to the thermal observables, the total free energy splits naturally into two distinct physical contributions:
\begin{equation}
\mathcal{F} = \mathcal{F}_{\text{bulk}} + \mathcal{F}_{\text{NL}} \; .
\end{equation}
The first term, $\mathcal{F}_{\text{bulk}}$, arises from the fractional square-root differential operator and represents the bulk-leakage contribution of the confined scalar field:
\begin{equation}\label{Fbulkdef}
\mathcal{F}_{\text{bulk}} = \frac{1}{2} \, \mathcal{F}_{\text{local}}(m) = \frac{T}{4} \sum_{n=-\infty}^\infty \int \frac{d^2\mathbf{p}}{(2\pi)^2} \ln\left( \omega_n^2 + \mathbf{p}^2 + m^2 \right) \; .
\end{equation}
Performing the standard Matsubara frequency summation via contour integration in the complex plane, this sum evaluates to:
\begin{equation}
\mathcal{F}_{\text{bulk}} = \int \frac{d^2\mathbf{p}}{(2\pi)^2} \left[ \frac{\omega_p}{4} + \frac{T}{2} \ln\left( 1 - e^{-\beta\omega_p} \right) \right] \; ,
\end{equation}
where $\omega_p = \sqrt{\mathbf{p}^2 + m^2}$, and the first term inside the brackets is the zero-point vacuum energy ($T=0$). Discarding the vacuum energy, the purely thermal part of the bulk free energy is:
\begin{equation}\label{Fbulkth}
\mathcal{F}_{\text{bulk}}^{th} = \frac{T}{2} \int \frac{d^2\mathbf{p}}{(2\pi)^2} \ln\left( 1 - e^{-\beta\sqrt{\mathbf{p}^2 + m^2}} \right) \; .
\end{equation}
In two-dimensional polar coordinates, introducing the energy variable $E = \sqrt{p^2 + m^2}$ with $p\,dp = E\,dE$, the integral transforms into:
\begin{equation}
\mathcal{F}_{\text{bulk}}^{th} = \frac{T}{4\pi} \int_m^\infty E \, \ln\left( 1 - e^{-\beta E} \right) dE \; .
\end{equation}
Expanding the logarithmic factor in the standard Bose series, $\ln(1-e^{-\beta E}) = -\sum_{n=1}^\infty \frac{e^{-n\beta E}}{n}$, and integrating term by term by parts, we find:
\begin{equation}\label{intEparts}
\int_m^\infty E \, e^{-n\beta E} \, dE = \frac{m}{n\beta} \, e^{-n\beta m} + \frac{e^{-n\beta m}}{(n\beta)^2} \; .
\end{equation}
Substituting (\ref{intEparts}) back into the series and using $\beta = 1/T$, we obtain the exact closed analytical expression for the bulk thermal free energy:
\begin{equation}\label{Fbulkres}
\mathcal{F}_{\text{bulk}}^{th} = -\frac{T^3}{4\pi} \sum_{n=1}^\infty e^{-n \frac{m}{T}} \left( \frac{m/T}{n^2} + \frac{1}{n^3} \right) \; .
\end{equation}
In the massless limit ($m \to 0$), Eq. (\ref{Fbulkres}) reduces to:
\begin{equation}\label{Fbulkm0}
\lim_{m \to 0} \mathcal{F}_{\text{bulk}}^{th} = -\frac{\zeta(3)}{4\pi} \, T^3 = -\frac{1}{2} \, \frac{\zeta(3)}{2\pi} \, T^3 \; ,
\end{equation}
where $\zeta(3) \simeq 1.202$ is Apéry's constant. The prefactor of $1/2$ relative to a standard local 2D scalar field confirms that the planar thermal bath carries an effective degree of freedom $g_{\rm eff} = 1/2$. This fractional degree of freedom is the exact thermodynamic manifestation of the kinematic confinement, reflecting the leakage of the field lines into the three-dimensional bulk.

The second term in Eq. (\ref{lnOE}), $\mathcal{F}_{\text{NL}}$, represents the purely non-local thermal contribution:
\begin{equation}\label{FNLdef}
\mathcal{F}_{\text{NL}} = -\frac{T}{2} \sum_{n=-\infty}^\infty \int \frac{d^2\mathbf{p}}{(2\pi)^2} \ln\left[ \mbox{erfc}\left( \frac{\sqrt{\omega_n^2 + \mathbf{p}^2 + m^2}}{\Lambda} \right) \right] \; .
\end{equation}
Using the spectral representation of functional determinants at finite temperature, the thermal part of $\mathcal{F}_{\text{NL}}$ is evaluated in terms of the phase shift $\delta_{\text{NL}}(E, \mathbf{p}) = -\mbox{Im}\ln[\mbox{erfc}(\sqrt{-E^2+\mathbf{p}^2+m^2+i\epsilon}/\Lambda)]$, with the same prescription for which the bulk factor $\ln\sqrt{-E^2+\omega_p^2+i\epsilon}$ has phase $\pi/2$ for $E>\omega_p$. As established in Section \ref{sec5}, for $E > \omega_p$, $\mbox{erfc}(+i\sqrt{E^2-\omega_p^2}/\Lambda) = 1 - i\,\mbox{erfi}(\sqrt{E^2-\omega_p^2}/\Lambda)$, where $\mbox{erfi}(z)$ is the imaginary error function. Hence, the non-local phase shift is:
\begin{equation}
\delta_{\text{NL}}(E, \mathbf{p}) = \mbox{arctan}\left[ \mbox{erfi}\left( \frac{\sqrt{E^2 - \omega_p^2}}{\Lambda} \right) \right] \; ,
\end{equation}
which yields the exact thermal non-local free energy:
\begin{eqnarray}\label{FNLexact}
\mathcal{F}_{\text{NL}}^{th} = -\frac{1}{\pi} \int \frac{d^2\mathbf{p}}{(2\pi)^2} \int_{\omega_p}^\infty \frac{dE}{e^{\beta E} - 1} \, 
\nonumber \\
\times \, \mbox{arctan}\left[ \mbox{erfi}\left( \frac{\sqrt{E^2 - \omega_p^2}}{\Lambda} \right) \right] \; .
\end{eqnarray}
Because $\mbox{erfi}(x) > 0$ for all $x > 0$, the phase shift is strictly positive, ensuring that $\mathcal{F}_{\text{NL}}^{th} \le 0$ for all temperatures. In the physical regime of interest, where the non-locality scale is much larger than the temperature ($T \ll \Lambda$), the thermal modes predominantly populate the low-energy regime $\sqrt{E^2-\omega_p^2} \ll \Lambda$. Expanding the imaginary error function to leading order, $\mbox{erfi}(x) \approx \frac{2}{\sqrt{\pi}} x$, and using $\mbox{arctan}(y) \approx y$, Eq. (\ref{FNLexact}) simplifies to:
\begin{equation}\label{FNLapprox}
\mathcal{F}_{\text{NL}}^{th} \approx -\frac{2}{\pi^{3/2} \Lambda} \int \frac{d^2\mathbf{p}}{(2\pi)^2} \int_{\omega_p}^\infty \frac{dE}{e^{\beta E} - 1} \, \sqrt{E^2 - \omega_p^2} \; .
\end{equation}
Exchanging the order of integration and performing the radial momentum integral over $\omega_p \in [m, E]$:
\begin{equation}
\int_m^E \omega_p \, \sqrt{E^2 - \omega_p^2} \, d\omega_p = \frac{1}{3} \left( E^2 - m^2 \right)^{3/2} \; ,
\end{equation}
Eq. (\ref{FNLapprox}) evaluates to:
\begin{equation}\label{FNLEintegral}
\mathcal{F}_{\text{NL}}^{th} \simeq -\frac{1}{3\pi^{5/2} \Lambda} \int_m^\infty dE \, \frac{\left( E^2 - m^2 \right)^{3/2}}{e^{\beta E} - 1} \; .
\end{equation}
We stress that Eq. (\ref{FNLEintegral}) is only the leading term of an expansion in $T/\Lambda$: the expansion of $\mbox{arctan}[\mbox{erfi}(x)]$ contains only odd powers of $x$, so the next correction is of relative order $T^2/\Lambda^2$ (i.e. $\mathcal{O}(T^6/\Lambda^3)$ in the free energy), and the approximation fails when $T \gtrsim \Lambda$. The exact result is always given by Eq. (\ref{FNLexact}).
Expanding the Bose factor $(e^{\beta E} - 1)^{-1} = \sum_{n=1}^\infty e^{-n\beta E}$ and using the integral representation of the modified Bessel function of the second kind,
\begin{equation}
\int_m^\infty \left( E^2 - m^2 \right)^{3/2} \, e^{-n\beta E} \, dE = \frac{3\,m^2}{(n\beta)^2} \, K_2(n\beta m) \; ,
\end{equation}
the non-local thermal free energy takes the explicit analytical form:
\begin{equation}\label{FNLseries}
\mathcal{F}_{\text{NL}}^{th} = -\frac{m^2 T^2}{\pi^{5/2} \Lambda} \, \sum_{n=1}^\infty \frac{K_2(n m/T)}{n^2} \; .
\end{equation}
In the regime $m \ll T \ll \Lambda$, the integral in Eq. (\ref{FNLEintegral}) can be performed directly with $m \to 0$ using that 
\begin{equation}
\int_0^\infty \frac{E^3 dE}{e^{\beta E}-1} = \frac{\pi^4}{15} T^4 \; ,    
\end{equation}
we obtain
\begin{equation}\label{FNLm0}
\mathcal{F}_{\text{NL}}^{th} \simeq -\frac{\pi^{3/2}}{45} \, \frac{T^4}{\Lambda} \; .
\end{equation}

From the thermal free energy $\mathcal{F}_{th} = \mathcal{F}_{\text{bulk}}^{th} + \mathcal{F}_{\text{NL}}^{th}$, the macroscopic thermodynamic pressure $P(T) = -\mathcal{F}_{th}$ exerted by the planar thermal bath is:
\begin{eqnarray}\label{PressureFinal}
P(T) &=& \frac{T^3}{4\pi} \sum_{n=1}^\infty e^{-n \frac{m}{T}} \left( \frac{m/T}{n^2} + \frac{1}{n^3} \right) 
\nonumber \\
&&
+ \frac{m^2 T^2}{\pi^{5/2} \Lambda} \, \sum_{n=1}^\infty \frac{K_2(n m/T)}{n^2} \; .
\end{eqnarray}
In the regime $m \ll T \ll \Lambda$, the equation of state takes the compact form:
\begin{equation}\label{PressureM0}
P(T) = \frac{\zeta(3)}{4\pi} \, T^3 + \frac{\pi^{3/2}}{45} \, \frac{T^4}{\Lambda} + \mathcal{O}\left( \frac{T^6}{\Lambda^3} \right) \; .
\end{equation}
Equations (\ref{PressureFinal}) and (\ref{PressureM0}) reveal essential features of the non-local planar dynamics:
\begin{enumerate}
    \item Both the bulk and non-local terms contribute positively to the total pressure ($P(T) > 0$).
    \item For $m \ll T \ll \Lambda$ the leading term reproduces the expected Stefan-Boltzmann $T^3$ scaling for a planar system, with half the standard single-scalar amplitude ($g_{\rm eff} = 1/2$) due to the bulk leakage.
    \item The non-local scale $\Lambda$ generates an outward thermal pressure correction proportional to $T^4/\Lambda$ in the regime $T \ll \Lambda$.
    \item Unlike truncated or rational models, there are no unphysical ghost poles, negative pressures, or mechanical instabilities at low temperatures ($T \to 0$). Macroscopic mechanical stability holds for the exact expression (\ref{FNLexact}), which is positive for all $T>0$, and not only for its expansion.
\end{enumerate}
For $T \gtrsim \Lambda$ the linearized result no longer applies and one must use Eq. (\ref{FNLexact}). Since $\mbox{arctan}[\mbox{erfi}(x)] \to \pi/2$ for $x \gg 1$, the non-local term saturates when $T \gg (\Lambda , m)$, $\mathcal{F}_{\text{NL}}^{th} \to \frac{1}{2}\mathcal{F}_{\text{local}}^{th}$, so that
\begin{equation}\label{PressureHighT}
\Delta P(T) \to \frac{\zeta(3)}{4\pi}\, T^3 \;\; , \;\; P(T) \to \frac{\zeta(3)}{2\pi}\, T^3 \; ,
\end{equation}
{\it i. e.}, the pressure of a full local $1+2$ scalar gas ($g_{\rm eff}=1$). Hence $g_{\rm eff}=1/2$ characterizes the window $m \ll T \ll \Lambda$, and the planar bath crosses over to $g_{\rm eff}=1$ at $T \sim \Lambda$, where the non-local mode is fully excited. We have confirmed numerically that $\Delta P/[\zeta(3)T^3/4\pi] \simeq 0.998$ at $T=30\,\Lambda$. The pressure remains strictly positive in all regimes.

Furthermore, the macroscopic entropy of the planar bath, defined as $S(T) = -\partial\mathcal{F}_{th}/\partial T = \partial P/\partial T$, is:
\begin{equation}\label{EntropyFinal}
S(T) = S_{\text{bulk}}(T) + \Delta S(T) \; ,
\end{equation}
where the bulk entropy is given by:
\begin{equation}\label{Sbulk}
S_{\text{bulk}}(T) = \frac{T^2}{4\pi} \sum_{n=1}^\infty e^{-n \frac{m}{T}} \left[ \frac{3}{n^3} + \frac{3 \left(\frac{m}{T}\right)}{n^2} + \frac{\left(\frac{m}{T}\right)^2}{n} \right] \; ,
\end{equation}
and the non-local correction in the regime $m \ll T \ll \Lambda$ reads:
\begin{equation}\label{Sm0}
\Delta S(T) = \frac{4\pi^{3/2}}{45} \, \frac{T^3}{\Lambda} \; .
\end{equation}
The entropy is strictly positive ($S(T) > 0$) for all temperatures, vanishing smoothly in the zero-temperature limit ($S \to 0$ as $T \to 0$) in accordance with the Third Law of Thermodynamics.

To provide a complete thermodynamic characterization of the planar system and rigorously confirm its thermal stability, we now evaluate the heat capacity (specific heat at constant area) of the thermal bath:
\begin{eqnarray}\label{CVdef}
C_V(T) &=& T \left( \frac{\partial S}{\partial T} \right)_V = - T \left( \frac{\partial^2 \mathcal{F}_{th}}{\partial T^2} \right)_V 
\nonumber \\
&=& C_{V,\text{bulk}}(T) + \Delta C_V(T) \; .
\end{eqnarray}
Differentiating the bulk entropy (\ref{Sbulk}) with respect to temperature (see the detailed derivation in the Appendix), we obtain the exact closed analytical expression for the bulk heat capacity:
\begin{eqnarray}\label{CVbulk}
C_{V,\text{bulk}}(T) = \frac{T^2}{4\pi} \sum_{n=1}^\infty e^{-n \frac{m}{T}} \left[ \frac{6}{n^3} + \frac{6 \left(\frac{m}{T}\right)}{n^2} 
\right.
\nonumber \\
\left.
+ \frac{3 \left(\frac{m}{T}\right)^2}{n} + \left(\frac{m}{T}\right)^3 \right] \; .
\end{eqnarray}
In the massless limit ($m \to 0$), the exponential factor approaches unity and all mass-dependent terms vanish, yielding:
\begin{equation}\label{CVbulkm0}
\lim_{m \to 0} C_{V,\text{bulk}}(T) = \frac{3\,\zeta(3)}{2\pi} \, T^2 = \frac{1}{2} \, C_{V,2D}^{\text{local}}(T) \; ,
\end{equation}
where $C_{V,2D}^{\text{local}}(T) = \frac{3\zeta(3)}{\pi} T^2$ is the standard heat capacity of a local massless scalar field in $1+2$ dimensions. The factor of $1/2$ provides a third independent thermodynamic confirmation—alongside the pressure (\ref{Fbulkm0}) and entropy (\ref{Sbulk})—that the planar thermal bath carries an effective fractional degree of freedom $g_{\rm eff} = 1/2$.

The non-local correction to the heat capacity, obtained by differentiating the non-local entropy in the regime $m \ll T \ll \Lambda$, is:
\begin{equation}\label{CVNLm0}
\Delta C_V(T) = T \, \frac{\partial \Delta S}{\partial T} = \frac{4\pi^{3/2}}{15} \, \frac{T^3}{\Lambda} + \mathcal{O}\left( \frac{T^5}{\Lambda^3} \right) \; .
\end{equation}
For finite mass $m$, the exact non-local correction can be expressed through the series of modified Bessel functions:
\begin{equation}\label{CVNLseries}
\Delta C_V(T) = \frac{m^4}{2\pi^{5/2} \Lambda T} \sum_{n=1}^\infty \left[ K_2\left(\frac{nm}{T}\right) + K_4\left(\frac{nm}{T}\right) \right] \; .
\end{equation}
Combining both contributions, the total heat capacity in the regime $m \ll T \ll \Lambda$ behaves as:
\begin{equation}\label{CVtotal}
C_V(T) = \frac{3\zeta(3)}{2\pi} \, T^2 + \frac{4\pi^{3/2}}{15} \, \frac{T^3}{\Lambda} + \mathcal{O}\left( \frac{T^5}{\Lambda^3} \right) \; .
\end{equation}
Equations (\ref{CVbulk}) and (\ref{CVtotal}) establish essential physical features of the planar thermal model:
\begin{enumerate}
    \item The heat capacity is strictly positive ($C_V(T) > 0$) for all temperatures $T > 0$, rigorously guaranteeing thermal stability against energy fluctuations ($\langle (\Delta E)^2 \rangle = T^2 C_V > 0$) and ruling out any negative-heat-capacity instability.
    \item In the low-temperature limit ($T \ll m$), $C_V(T)$ is exponentially suppressed by the mass gap, $C_V(T) \sim \frac{m^3}{4\pi T} e^{-m/T} > 0$, smoothly approaching zero as $T \to 0$ in strict compliance with the Third Law of Thermodynamics.
    \item For $m \ll T \ll \Lambda$, the non-local scale generates a positive enhancement $\propto T^3/\Lambda$; for $T \gg \Lambda$ the heat capacity saturates at the local $1+2$ value, $C_V \to 3\zeta(3)T^2/\pi$ ($g_{\rm eff}=1$), consistently with Eq. (\ref{PressureHighT}).
\end{enumerate}

\begin{figure}[htbp]
    \centering
    \includegraphics[width=1\columnwidth]{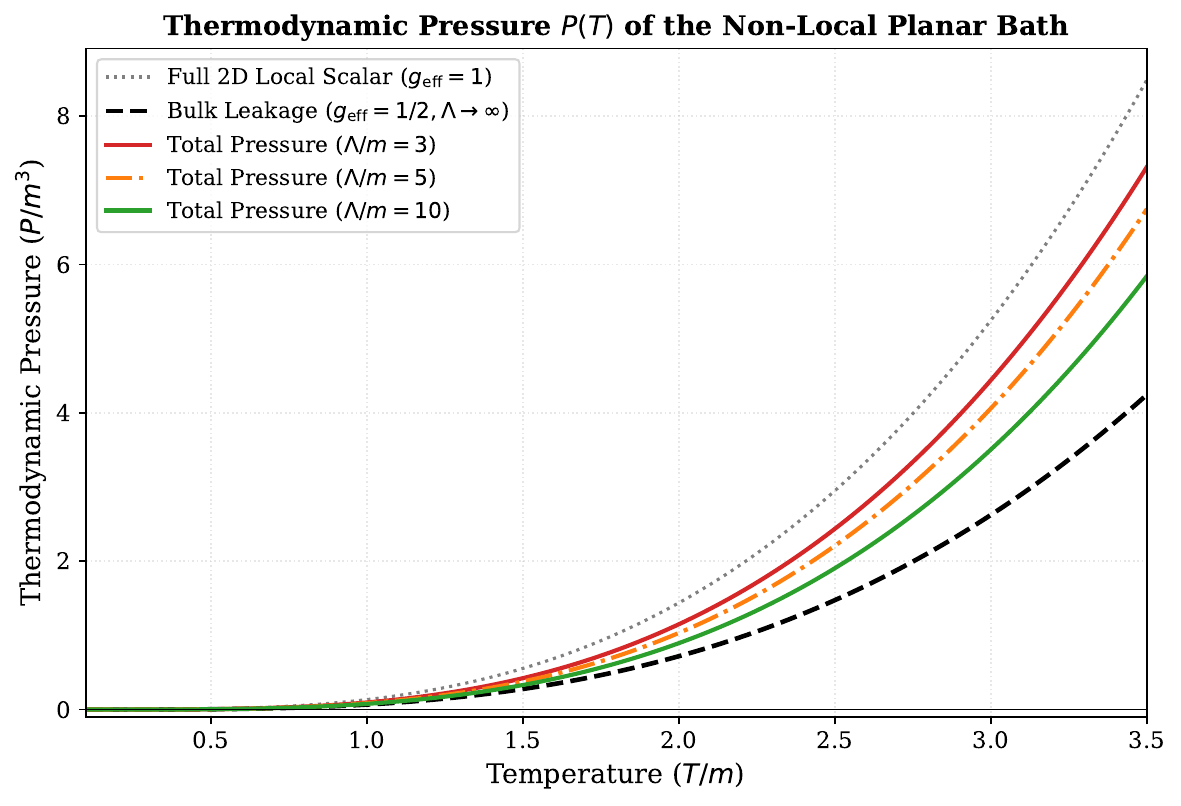}
    \caption{The thermodynamic pressure $P(T)$ of the planar thermal bath as a function of $T/m$. The dashed black curve represents the local bulk-leakage limit ($g_{\rm eff}=1/2, \Lambda\to\infty$), while the solid colored curves show the total pressure for $\Lambda/m = 3, 5$, and $10$. For comparison, the dotted gray curve displays the full 2D local scalar gas ($g_{\rm eff}=1$). The non-local term is evaluated exactly from Eq. (\ref{FNLexact}). All curves are positive definite and lie between the $g_{\rm eff}=1/2$ and $g_{\rm eff}=1$ limits, confirming macroscopic thermodynamic stability.}
    \label{fig:pressure}
\end{figure}

Figure \ref{fig:pressure} illustrates the thermodynamic pressure $P(T)$ as a function of the temperature for different values of $\Lambda/m$, computed with the exact non-local contribution (\ref{FNLexact}) (no expansion in $T/\Lambda$), compared to the bulk-leakage limit ($\Lambda \to \infty$) and the full 2D local scalar field ($g_{\rm eff}=1$).

\begin{figure}[htbp]
    \centering
    \includegraphics[width=1\columnwidth]{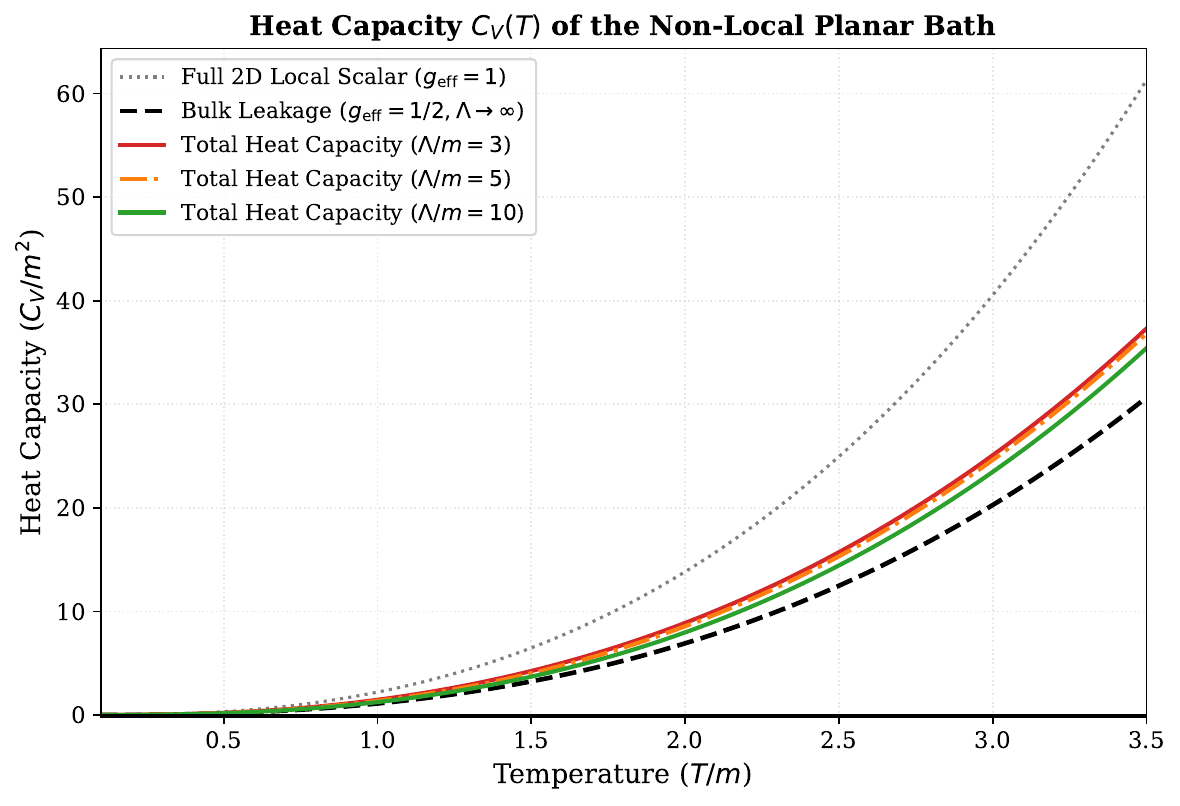}
    \caption{The heat capacity $C_V(T)$ of the planar thermal bath as a function of $T/m$. The dashed black curve denotes the local bulk-leakage limit ($g_{\rm eff}=1/2, \Lambda\to\infty$), while the solid colored curves display the total heat capacity for $\Lambda/m = 3, 5$, and $10$. For comparison, the dotted gray curve illustrates the full 2D local scalar gas ($g_{\rm eff}=1$). The non-local term is evaluated exactly from Eq. (\ref{FNLexact}). All curves are strictly positive and vanish smoothly as $T \to 0$, corroborating thermal stability and the Third Law of Thermodynamics.}
    \label{fig:heat_capacity}
\end{figure}

Similarly, Figure \ref{fig:heat_capacity} displays the thermal heat capacity $C_V(T)=T\,\partial^2 P/\partial T^2$ as a function of $T/m$ for different values of $\Lambda/m$, again with the exact non-local contribution, compared to the local bulk-leakage limit ($g_{\rm eff}=1/2$) and the full 2D local scalar gas ($g_{\rm eff}=1$). The profiles confirm the strictly positive response across the entire thermal domain, the exponential damping at low temperatures ($T \ll m$) in full agreement with the Third Law of Thermodynamics, and the positive enhancement $\sim T^3/\Lambda$ induced by the non-local form factor for $m \ll T \ll \Lambda$, which saturates towards the $g_{\rm eff}=1$ curve for $T \gg \Lambda$.

\section{Conclusions}

In this paper, we have investigated the dimensional reduction from $1+3$ to $1+2$ dimensions of a non-local scalar field model governed by an entire-function form factor $e^{-\Box_E/\Lambda^2}$. By constraining external sources to a bidimensional plane, we derived the exact planar Green's function, expressed in terms of the complementary error function, and formulated the corresponding effective non-local pseudo-differential planar Lagrangian.

At zero temperature, we determined the static interaction energy between two point-like charges. We proved that the non-local scale $\Lambda$ acts as an intrinsic ultraviolet regulator that removes the classical $1/r$ singularity at the origin, yielding a finite value $U(0)$, while smoothly recovering the standard 3D Yukawa potential at long distances without introducing spurious ghost poles or negative-energy barriers.

Furthermore, we established the macroscopic causality of the planar model by evaluating the retarded, advanced, and Feynman Green's functions in coordinate space. We showed that the field commutator and causal propagators vanish identically outside the light-cone ($\bar{x}^2 < 0$), strictly confining information propagation within the light-cone. The Feynman Green's function remains regular at the light-cone coincidence limit, confirming the UV softening of quantum fluctuations. Moreover, we verified the unitarity of the planar model at tree level using the optical theorem.

Finally, we analyzed the finite-temperature properties of the effective planar model using the Matsubara formalism. The exact thermal free energy completely resolves the thermodynamic stability issues of earlier truncated models. The equation of state reveals an effective degree of freedom $g_{\rm eff} = 1/2$ in the window $m \ll T \ll \Lambda$, providing thermodynamic confirmation of kinematic bulk leakage, and crosses over to $g_{\rm eff}=1$ for $T \gg \Lambda$. Both the pressure, entropy, and heat capacity are strictly positive for all temperatures, establishing unconditional macroscopic thermodynamic stability and demonstrating that the entire-function non-locality regularizes short-distance behavior without spoiling quantum or thermodynamic consistency.

These results establish a consistent quantitative foundation for non-local planar field theories. As a natural outlook, the non-local planar framework developed here can be extended to include self-interactions (such as $\lambda\phi^4$) and gauge fields, enabling the study of dynamical symmetry breaking, vacuum condensates, and phase transitions in planar systems relevant to low-dimensional condensed matter physics.

\section*{Appendix: Derivation of the Exact Thermal Entropy and Heat Capacity}

\subsection*{Derivation of the Bulk Entropy}

To rigorously derive the bulk entropy expression presented in Eq. (\ref{Sbulk}), we evaluate the temperature derivative of the bulk thermal free energy (\ref{Fbulkres}):
\begin{equation}
    \mathcal{F}_{\text{bulk}}^{th}(T) = -\frac{1}{4\pi} \sum_{n=1}^{\infty} e^{-n \frac{m}{T}} \left( \frac{m \, T^2}{n^2} + \frac{T^3}{n^3} \right) \; .
\end{equation}
The corresponding entropy is defined by $S_{\text{bulk}} = -\partial \mathcal{F}_{\text{bulk}}^{th} / \partial T$. Applying the product rule to each term inside the series, we define $A(T) = e^{-n \, m/T}$ and $B(T) = m \, T^2/n^2 + T^3/n^3$. Their respective derivatives with respect to temperature are:
\begin{equation}
    \frac{\partial A}{\partial T} = e^{-n \frac{m}{T}} \left( \frac{n\,m}{T^2} \right) \quad \text{and} \quad \frac{\partial B}{\partial T} = \frac{2mT}{n^2} + \frac{3T^2}{n^3} \; .
\end{equation}
The product rule yields:
\begin{subequations}
\begin{eqnarray}
\frac{\partial A}{\partial T} \, B(T) &=& e^{-n \frac{m}{T}} \left( \frac{m^2}{n} + \frac{m T}{n^2} \right) \; , 
\\
A(T) \, \frac{\partial B}{\partial T} &=& e^{-n \frac{m}{T}} \left( \frac{2 m T}{n^2} + \frac{3 T^2}{n^3} \right) \; .
\end{eqnarray}
\end{subequations}
Summing these contributions and factoring out $T^2$, we obtain the exact analytical form for the entropy of the bulk mode in the planar thermal bath:
\begin{equation}
S_{\text{bulk}}(T) = \frac{T^2}{4\pi} \, \sum_{n=1}^{\infty} e^{-n \frac{m}{T}} \left[ \frac{3}{n^3} + \frac{3 \left(\frac{m}{T}\right)}{n^2} + \frac{\left(\frac{m}{T}\right)^2}{n} \right] \; .
\end{equation}

As a physical consistency check, evaluating the massless limit ($m \to 0$) of this result, the exponential factor approaches unity and all mass-dependent terms inside the brackets vanish, leaving only the inverse cube term:
\begin{equation}
S_{\text{bulk}}^{m \to 0}(T) = \frac{3\, T^2}{4\pi} \, \sum_{n=1}^{\infty} \frac{1}{n^3} = \frac{3 \, \zeta(3)}{4\pi} \, T^2 \; ,
\end{equation}
where $\zeta(3) \simeq 1.202$ is the Riemann zeta function. This result recovers exactly half of the standard thermodynamic entropy of a free massless bosonic gas in $1+2$ dimensions ($S_{2D} = \frac{3\zeta(3)}{2\pi} T^2$), confirming the fractional degree of freedom $g_{\rm eff} = 1/2$ and validating the thermodynamic consistency of the planar dimensional reduction.

\subsection*{Derivation of the Bulk Heat Capacity}

To rigorously derive the exact bulk heat capacity presented in Eq. (\ref{CVbulk}), we evaluate the temperature derivative of the bulk entropy (\ref{Sbulk}):
\begin{eqnarray}
&&
C_{V,\text{bulk}}(T) = T \, \frac{\partial S_{\text{bulk}}}{\partial T} 
\nonumber \\
&&
= \frac{T}{4\pi} \sum_{n=1}^\infty \frac{\partial}{\partial T} \left[ e^{-n\frac{m}{T}} \left( \frac{3T^2}{n^3} + \frac{3mT}{n^2} + \frac{m^2}{n} \right) \right] \, . \;\;\;
\end{eqnarray}
Applying the product rule to each term in the series, we define $A(T) = e^{-n m/T}$ and $C(T) = 3T^2/n^3 + 3mT/n^2 + m^2/n$. Their derivatives with respect to temperature are:
\begin{equation}
\frac{\partial A}{\partial T} = e^{-n\frac{m}{T}} \left( \frac{nm}{T^2} \right) \quad \text{and} \quad \frac{\partial C}{\partial T} = \frac{6T}{n^3} + \frac{3m}{n^2} \; .
\end{equation}
The corresponding product contributions evaluate to:
\begin{subequations}
\begin{eqnarray}
\frac{\partial A}{\partial T} \, C(T) &=& e^{-n\frac{m}{T}} \left[ \frac{3m}{n^2} + \frac{3m^2}{nT} + \frac{m^3}{T^2} \right] \; , \\
A(T) \, \frac{\partial C}{\partial T} &=& e^{-n\frac{m}{T}} \left[ \frac{6T}{n^3} + \frac{3m}{n^2} \right] \; .
\end{eqnarray}
\end{subequations}
Summing these contributions yields:
\begin{equation}
\frac{\partial S_{\text{bulk}, n}}{\partial T} = \frac{e^{-n\frac{m}{T}}}{4\pi} \left[ \frac{6T}{n^3} + \frac{6m}{n^2} + \frac{3m^2}{nT} + \frac{m^3}{T^2} \right] \; .
\end{equation}
Multiplying by $T$ and factoring out $T^2$, we arrive at the closed analytical result:
\begin{equation}
C_{V,\text{bulk}}(T) = \frac{T^2}{4\pi} \sum_{n=1}^\infty e^{-n\frac{m}{T}} \left[ \frac{6}{n^3} + \frac{6\left(\frac{m}{T}\right)}{n^2} + \frac{3\left(\frac{m}{T}\right)^2}{n} + \left(\frac{m}{T}\right)^3 \right] \; ,
\end{equation}
which strictly confirms Eq. (\ref{CVbulk}). Taking the massless limit ($m \to 0$):
\begin{equation}
\lim_{m \to 0} C_{V,\text{bulk}}(T) = \frac{6\,T^2}{4\pi} \sum_{n=1}^\infty \frac{1}{n^3} = \frac{3\,\zeta(3)}{2\pi} \, T^2 \; ,
\end{equation}
in full agreement with Eq. (\ref{CVbulkm0}).

%


%

\end{document}